\documentclass[12pt]{article}
\usepackage{amsmath,amssymb}
\usepackage{hyperref}
\usepackage{authblk}
\usepackage{geometry}
\title{\textbf{Digamma-Function Representation of the Ground-State Energy in Antiferromagnetic Heisenberg $\mathrm{XXX}_s$ Spin Chains}}
\author{Reaz Shafqat}
\affil{Department of Mathematics and Natural Sciences, BRAC University, Dhaka, Bangladesh}
\date{December 2025}
\begin{document}
\maketitle
\begin{abstract}
The antiferromagnetic Heisenberg spin chain remains a central framework for exploring exactly solvable models within quantum integrable systems. For the isotropic XXX chain, the ground-state energy per site of the spin-$\frac{1}{2}$ system is famously given by $\ln{2}$. Extending the classic formulations to arbitrary spin-s, Takhtajan and Babujian derived two separate finite-series expressions for integer and half-integer spin representations. Current work introduces a unified analytical expression for the ground-state energy density in terms of the digamma function. This compact formulation reproduces both the Takhtajan Babujian series.
\end{abstract}
\section*{Introduction\footnote{This work is based on the master's thesis of Reaz Shfqat submitted to the Department of Theoretical Physics of the University of Dhaka, Session: 21-22.}}
The study of quantum spin chains occupies a central role in condensed--matter physics, integrable models, and mathematical physics. Among these systems, the antiferromagnetic Heisenberg model stands out as a paradigmatic exactly solvable many-body model, capturing key features of strong correlations, quantum criticality, and universal behaviour in one dimension. Since Bethe’s original solution of the spin-\(\tfrac12\) XXX chain \cite{Bethe1931}, the model has remained a cornerstone in the development of integrable systems, inspiring powerful frameworks such as the coordinate and Algebraic Bethe Ansatz \cite{yang1969,korepin1993,klumper1998}, the quantum inverse scattering method \cite{reshetikhin1983}, and deep connections to quantum groups and conformal field theory (CFT) \cite{haldane1983,haldane1983b,ZamolodchikovZamolodchikov1979,Affleck1986}.  

For higher-spin generalisations of the isotropic XXX chain, the exact analytic structure becomes richer. Independently, Takhtajan \cite{Takhtajan1982} and Babujian \cite{Babujian1983} constructed closed-form thermodynamic Bethe–Ansatz (TBA) expressions for the ground-state energy density for arbitrary spin \(s\). Their analyses established the now-famous distinction between integer and half-integer spin chains, later connected to the emergence of topological \(\theta\)-terms in the low-energy nonlinear sigma model (NLSM) description \cite{haldane1983,haldane1983b,ZhangSchulzZiman1989}. These results revealed that the ground-state energy admits two finite-series expansions that differ structurally for integer and half-integer \(s\), a reflection of the fundamental dichotomy between gapped (integer) and critical (half-integer) chains predicted by the Haldane conjecture\cite{haldane1983,haldane1983b}.  

Despite the elegance and completeness of the Takhtajan–Babujian formalism, the resulting expressions do not provide a single, analytically continuous formula valid for all real \(s\). Moreover, modern developments—such as anyonic spin chains, fractionalised excitations, and extensions via quantum-group symmetry \cite{Feiguin2007,Lecheminant2015,Mourigal2013,Chen2015}—motivate a unified perspective on the ground-state energy that treats \(s\) as a continuous variable. This work proposes a compact analytic representation of the antiferromagnetic vacuum energy density in terms of the digamma functions. This formulation reproduces the Takhtajan–Babujian results exactly. The proposed expression offers an analytically tractable, physically transparent, and computationally convenient alternative to traditional finite-series expansions, thereby simplifying both theoretical analysis and numerical evaluation.

The remainder of this paper is organised as follows. The following section reviews the essential theoretical foundations of the Heisenberg XXX chain: the Takhtajan–Babujian generalisation to arbitrary spin-s, and the resulting integer/half-integer series structure for the ground-state energy density. Afterwards, the digamma-function representation of the ground state energy density with a detailed derivation is outlined.

\section*{Babujian’s thermodynamic Bethe ansatz for $\mathrm{XXX}_s$ chain}
A brief overview of Babujian’s thermodynamic Bethe ansatz (TBA) for the antiferromagnetic spin-$s$ Heisenberg XXX chain is presented, where $s$ may take either integer or half-integer values~\cite{babujian1982,Babujian1983}. This formulation generalizes the earlier spin-$\tfrac12$ treatment developed by Takahashi and Yang--Yang~\cite{takahashi1971,yang1969}. The analysis starts from the Bethe equations for $N$ sites, $M$ down spins (magnons) with periodic boundary conditions. The rapidities shift as $\lambda_j\to i\lambda_j+\tfrac12$, one obtains\begin{equation}
\left(\frac{\lambda_j-is}{\lambda_j+is}\right)^N
 = \prod_{\substack{\ell=1\\ \ell\neq j}}^{M}
   \frac{\lambda_j-\lambda_\ell-i}{\lambda_j-\lambda_\ell+i} .
\label{eq:spin-s-bethe}
\end{equation}
In the thermodynamic limit, solutions are assumed to form ``strings’’ of
length $n$ in the complex plane,
\begin{equation}
\lambda_j^{n,\alpha}
 = \lambda_j^{n} + i\,(n+1-2\alpha),\qquad
\alpha=1,\dots,n ,
\end{equation}
with real string centres $\lambda_j^n$. The corresponding Bethe ansatz equation will take on the following form:
\begin{equation}
\left( \frac{\lambda_j + i s}{\lambda_j - i s} \right)^{N}
=
\prod_{\substack{k=1 \\ k \neq j}}^{\xi_1}
    \frac{\lambda_j - \lambda_k + i}{\lambda_j + \lambda_k - i}
\;
\prod_{\substack{k=1 \\ k \neq j}}^{\xi_2}
    \frac{\lambda_j - \lambda_k + 2 i}{\lambda_j - \lambda_k - 2 i}
\cdots
\prod_{\substack{k=1 \\ k \neq j}}^{\xi_{2s}}
    \frac{\lambda_j - \lambda_k + 2 s i}{\lambda_j - \lambda_k - 2 s i} .
    \label{eq:verylong bethe}
\end{equation}
If $\xi_n$ denotes the number
of $n$–strings, then
\begin{equation}
M = \sum_{n=1}^\infty n\,\xi_n .
\end{equation}

Multiplying \eqref{eq:verylong bethe} over the members of a given string
and taking the logarithm yields Bethe equations for the string centres,
\begin{equation}
N\,\theta_{n,2s}(\lambda_j^n)
 = 2\pi J_j^n
   + \sum_{k=1}^\infty\sum_{\ell=1}^{\xi_k}
     \Xi_{n,k}(\lambda_j^n-\lambda_\ell^k),
\label{eq:string-centre-bethe}
\end{equation}
where $J_j^n$ are integer or half–integer quantum numbers and
\begin{equation}
\theta_n(\lambda) = 2\arctan\!\frac{\lambda}{n},\qquad
\theta_{m,p}(\lambda)
 = \sum_{\ell=1}^{\min(m,p)}\theta_{m+p+1-2\ell}(\lambda)
\end{equation}
are basic scattering phases. The $\Xi_{m,k}$ are fixed linear
combinations of the $\theta_\ell$ that encode scattering between
$m$– and $k$–strings. The multi–string scattering phase can be expressed through the kernels
$\Xi_{m,p}(\lambda)$ using the convolution identity
\begin{equation}
\theta_{m,p}(\lambda)
 = \hat{p}\!\left(\Xi_{m,p}(\lambda)
   + \pi\,\mathrm{sgn}\,\lambda\,\delta_{m,p}\right),
\label{eq:theta-mp}
\end{equation}
where
\begin{equation}
\hat{p}g(\lambda)=\int_{-\infty}^{\infty}
p(\lambda-\lambda')\,g(\lambda')\,d\lambda',
\qquad
p(\lambda)=\left(4\cosh\frac{\pi\lambda}{2}\right)^{-1}.
\label{eq:p-kernel}
\end{equation}

In the thermodynamic limit, the discrete roots are described by
continuous densities of occupied and unoccupied quantum numbers,
$\rho_n(\lambda)$ and $\tilde{\rho}_n(\lambda)$, defined such that
$\rho_n(\lambda)\,d\lambda$ gives the number of $n$–strings with
centres in $[\lambda,\lambda+d\lambda]$ (and similarly for
$\tilde{\rho}_n$).  Differentiating
\eqref{eq:string-centre-bethe} and converting sums into integrals, one
obtains the basic linear relation
\begin{equation}
2\pi\bigl[\rho_n(\lambda)+\tilde{\rho}_n(\lambda)\bigr]
 = \theta'_{n,2s}(\lambda)
   - \sum_{k=1}^\infty
      \int_{-\infty}^{\infty}
         K_{n,k}(\lambda-\lambda')\,
         \rho_k(\lambda')\,d\lambda',
\label{eq:rho-eq-brief}
\end{equation}

where $K_{n,k}(\lambda)=\frac{d}{d\lambda}\Xi_{n,k}(\lambda)/2\pi$ is
the TBA kernel. In the thermodynamic limit, the relation between particle and hole densities
takes the form
\begin{equation}
\tilde{\rho}_m(\lambda)
 = -\sum_{n=1}^{\infty}
   \,\hat{A}_{m,n}\!\left[\rho_n(\lambda)
      - p(\lambda)\,\delta_{n,2s}\right],
\label{eq:rhohat}
\end{equation}
where $\hat{A}_{m,n}$ denotes the integral operator built from the kernel
$A_{m,n}(\lambda)$.  
The energy  per site can be expressed
directly in terms of the $\rho_n$,
\begin{equation}
\frac{E_s}{N}
 = -\frac12\sum_{n=1}^\infty
     \int \theta'_{n,2s}(\lambda)\,\rho_n(\lambda)\,d\lambda
     \label{eqn: energy of bethe}
\end{equation}

The equilibrium distributions at temperature $T$ and magnetic field $H$
follow from minimizing the free energy $F=E_s-TS-HS^z$ at $T=0 \text{ and } H=0$ with respect to $\rho_n\text{ and } \tilde{\rho}_n$, where the
Entropy has the standard TBA form
\cite{yang1969,takahashi1999}.  Introducing the pseudoenergies
\begin{equation}
\varepsilon_n(\lambda)
 = T\ln\frac{\tilde{\rho}_n(\lambda)}{\rho_n(\lambda)},
\end{equation}
The stationarity condition yields coupled non–linear integral equations
of the form
\begin{equation}
\varepsilon_n(\lambda)
 = Hn - \tfrac12\theta'_{n,2s}(\lambda)
   - T \sum_{k=1}^\infty
        (K_{n,k} * \ln[1+e^{-\varepsilon_k/T}])(\lambda),
\label{eq:TBA-main}
\end{equation}
with $*$ denoting convolution.  These are Babujian’s TBA equations for the spin-$s$ chain. Finally, the zero–temperature limit $T\to0$ of \eqref{eq:TBA-main} determines the ground state. One finds that only strings of length $2s$ are occupied in the antiferromagnetic vacuum, with pseudoenergy
\begin{equation}
\varepsilon_{2s}(\lambda)
 = -\frac{\pi}{4\cosh(\frac{\pi}{2}\lambda)},
\qquad
\varepsilon_{n\neq2s}(\lambda)\ge0 .
\end{equation}
Substituting this into the expression for the free energy gives the
ground–state energy
\begin{equation}
\frac{E_s^0}{N}
 = - \sum_{k=1}^{2s}\beta(k),
\qquad
\beta(k)=\int_0^1\frac{x^{k-1}}{1+x}\,dx .
\label{eq:beta}
\end{equation}

The integral in \eqref{eq:beta} is evaluated by using the recursive reduction of power in the numerator. So for integer and half-integer, one gets two separate reduction formulas. Thus, evaluating the sum, Babujian obtains the compact formulas
\begin{align}
\text{integer } s:\quad
\frac{E_s^0}{N} &= -\sum_{k=1}^{s}\frac{1}{2k-1},\\[0.3ex]
\text{half-integer } s:\quad
\frac{E_s^0}{N} &= -\left(\ln2+\sum_{k=1}^{s-\frac12}\frac{1}{2k}\right),
\label{eq: babujan energy}
\end{align}

which display the characteristic difference between integer and
half–integer spins, in line with Haldane’s conjecture on the low-energy
physics of Heisenberg chains \cite{haldane1983,haldane1983b}.
\section*{Proposed Expression}
Following \cite{Doikou1997,takahashi1971}, it can simply be stated that the ground state of the antiferromagnetic Heisenberg spin chain contains no holes. Hence in \eqref{eq:rhohat} $\tilde{\rho}_{m}(\lambda) =0$. So, 
\[\rho(\lambda) = p(\lambda) \delta_{n,2s} = \frac{-\delta_{n,2s}}{4 \cosh(\pi\lambda/2)} \]

The above equation suggests that the ground state only contains strings of length: $2s$. Now, in order to determine the ground state energy \eqref{eqn: energy of bethe} is considered. Since the ground state contains only strings of length $2s$, the choice $n = 2s$ follows directly. Hence,
\begin{equation*}
\theta_{2s,2s}(\lambda)
 = \sum_{\ell=1}^{2s}\theta_{4s+1-2\ell}(\lambda)  = 2 \sum_{\ell=1}^{2s} \arctan\left(\frac{\lambda}{4s+1-2l}\right)
\end{equation*}
Now differentiating w.r.t $\lambda$:
\begin{equation*}
\theta_{2s,2s}'(\lambda)
  \sum_{\ell=1}^{2s} \frac{4s+1-2l}{\lambda^2+(4s+1-2l)^2}
\end{equation*}
Plugging the expression of $\rho(\lambda)$ and $\theta_{2s,2s}'(\lambda)$ in \eqref{eqn: energy of bethe}, giving:  
\begin{equation}
\frac{E(s)}{N}=\sum_{l = 1}^{2s}\frac{-1}{4}\int_{-\infty}^{\infty}
\frac{4s+1-2l}{\lambda^{2}+(4s+1-2l)^{2}}
\frac{d\lambda}{\cosh(\pi\lambda/2)},
\label{eq:Idef}
\end{equation}
Integrals involving the hyperbolic secant kernel arise naturally in integrable spin chains and thermodynamic Bethe ansatz calculations. For $4s+1-2l>0$ with $0\leq l \leq 2s$ the standard Laplace representation is employed:
\begin{equation}
\frac{4s+1-2l}{\lambda^{2}+(4s+1-2l)^{2}}
=\int_{0}^{\infty} e^{-(4s+1-2l)t}\cos(\lambda t)\, dt.
\label{eq:laplace}
\end{equation}
Substituting \eqref{eq:laplace} into \eqref{eq:Idef} and interchanging the order of integration gives
\begin{align}
\frac{E(s)}{N}
&=\frac{-1}{4}\int_{0}^{\infty} e^{-(4s+1-2l)t}
\left(
\int_{-\infty}^{\infty}
\frac{\cos(\lambda t)}
     {\cosh(\pi\lambda/2)}
\, d\lambda
\right) dt \nonumber \\
&=\frac{-1}{4}\int_{0}^{\infty} e^{-(4s+1-2l)t}\, J(t)\, dt,
\label{eq:I-J}
\end{align}
where
\begin{equation}
J(t)=\int_{-\infty}^{\infty}
\frac{\cos(\lambda t)}
     {\cosh(\pi\lambda/2)}\, d\lambda.
\label{eq:Jdef}
\end{equation}
To compute $J(t)= \int_{-\infty}^{\infty}
\frac{\cos(\lambda t)}
     {\cosh(\pi\lambda/2)}\, d\lambda$, first this Fourier cosine transform is converted to the Fourier transform:
\begin{equation}
\int_{-\infty}^{\infty}
\frac{\cos(\lambda t)}
     {\cosh(\pi\lambda/2)}\, d\lambda= \mathcal{\Re}\left( \int_{-\infty}^{\infty}\frac{e^{-i\lambda t}}{(\cosh \pi \lambda /2)}\, d\lambda \right)
\label{eq:sech-transform}
\end{equation}
First, the following substitution has been made:
\[
\mu=\frac{\pi\lambda}{2},\qquad 
\lambda=\frac{2\mu}{\pi},\qquad 
d\lambda=\frac{2}{\pi}\, d\mu,
\]
to rewrite \eqref{eq:Jdef} as
\begin{equation}
J(t)=\frac{2}{\pi}\int_{-\infty}^{\infty}
\frac{\exp
\left(i\frac{2 t}{\pi}\mu\right)}{\cosh \mu}\, d\mu.
\end{equation}
Expressing $\cosh\left(\mu\right) = \frac{2}{e^\mu + e^{-\mu}}$:
\[
J(t)
= \frac{2}{\pi}\int_{-\infty}^{\infty}\frac{\exp
\left(i\frac{2 t}{\pi}\mu\right)}{e^\mu + e^{-\mu}}\, d\mu
\]
Introducing the substitution \(x = e^{\mu}\), which implies \(\mu = \ln x\), the integration limits are transformed accordingly:
\[
J(t)
= \frac{4}{\pi}\int_{-\infty}^{\infty}\frac{\exp
\left(i\frac{2 t}{\pi}\mu\right)}{e^\mu + e^{-\mu}}\, d\mu =  \frac{4}{\pi}\int_{0}^{\infty}\frac{x^
{\left(i\frac{2 t}{\pi}\right)}}{1 + x^2}\, dx = \frac{4}{\pi}\int_{0}^{\infty}\frac{x^{\left(i\frac{2 t}{\pi}+1\right)-1}}{1 + x^2}  dx\]
Setting \(\alpha = i\frac{2 t}{\pi}+1 \) and \(\beta = 2\) gives:
\[
J(t)
= \frac{4}{\pi}\int_{0}^{\infty}\frac{x^{
\alpha-1}}{1 + x^\beta} dx
\]
Now, letting \( y = (1 + x^{\beta})^{-1} \), since \( 0 < \Re \left(\alpha \right) < \beta \), the integral $J(t)$ becomes:
\[
\frac{4}{\pi}\int_{0}^{\infty} \frac{x^{\alpha - 1}}{1 + x^{\beta}}\, dx
= \frac{4}{\pi \beta} \int_{0}^{1} 
y^{-\alpha/\beta} (1 - y)^{\alpha/\beta - 1}\, dy
= \frac{4}{\pi\beta} B\!\left( 1 - \frac{\alpha}{\beta},\, \frac{\alpha}{\beta} \right)
\]
The expression reduces to the Beta function. Using its representation in terms of Gamma functions, one obtains:
\[
= \frac{4}{\pi\beta}
\, \Gamma\!\left( 1 - \frac{\alpha}{\beta} \right)
\, \Gamma\!\left( \frac{\alpha}{\beta} \right)
= \frac{4\pi}{\pi\beta} \csc\!\left( \frac{\alpha \pi}{\beta} \right).
\]
Applying values of \(\alpha \text{ \& }\beta \):
\begin{equation}
    J(t) = 2 \csc\{\left(i\frac{2 t}{\pi}+1 \right)\frac{\pi}{2}\} = \sec(i t) =2 \text{sech(t)}
    \label{eq:J-final}
\end{equation}
Substituting \eqref{eq:J-final} into \eqref{eq:I-J} gives the simpler representation
\begin{equation}
\frac{E(s)}{N}=\frac{-2}{4}\int_{0}^{\infty} e^{-(4s+1-2l)t} \text{sech(t)}\, dt.
\label{eq:I-sech}
\end{equation}
Use the standard expansion
\begin{equation}
\text{sech(t)} = \frac{2}{e^t+e^{-t}}=\frac{2e^-t}{1+e^{-2t}},
\label{eq:sech-series1}
\end{equation}
Now the integral interval in \eqref{eq:I-sech} is $0$ to $\infty$. On this interval $0<e^{-2t}\leq1$. So, the binomial expansion is taken into account \eqref{eq:sech-series1}.
\begin{equation}
\text{sech(t)} =\frac{2e^-t}{1+e^{-2t}}= 2\sum_{n=1}^{\infty} (-1)^n e^{-(2n+1)t}
\label{eq:sech-series2}
\end{equation}
which, when inserted into \eqref{eq:I-sech}, yields
\begin{align}
\frac{E(s)}{N}
&=\frac{-4}{4}\sum_{n=0}^{\infty}(-1)^n\int_{0}^{\infty}
e^{-(4s+1-2l+2n+1)t}\, dt \nonumber \\
&=-\sum_{n=0}^{\infty}\frac{(-1)^n}{4s+1-2l+2n+1}.
\label{eq:Iseries}
\end{align}
One defines:
\[
A=\frac{4s+1-2l+1}{2},
\qquad 4s+1-2l+2n+1=2(n+A),
\]
so that \eqref{eq:Iseries} becomes
\begin{equation}
\frac{E(s)}{N}=-\sum_{n=0}^{\infty}\frac{(-1)^n}{n+A}.
\label{eq:series-A}
\end{equation}
\eqref{eq:series-A} can be represented by Lerch transcendence, $\Phi(z,s,a) = \sum_{k=0}^{\infty} \frac{z^k}{(a+k)^s}$ by setting, $z=-1,s=1, k =n \text{ and } a=A$. Therefore, the following equation is obtained, 
\begin{equation}
    \Phi(-1,1,A) = \sum_{n=0}^{\infty} \frac{(-1)^n}{(A+n)}
    \label{eq:lerch trans}
\end{equation}
It is known that the Lerch transcendent\cite{GradshteynRyzhik2000} has a digamma function representation expressed as:
\begin{equation}
\sum_{n=0}^{\infty} \frac{(-1)^n}{n+z}
= \frac12\!\left[
\psi\!\left(\frac{z+1}{2}\right)
-
\psi\!\left(\frac{z}{2}\right)
\right],
\label{eq:summation-identity}
\end{equation}
where $\psi$ denotes the digamma function.  
From \eqref{eq:series-A} and \eqref{eq:summation-identity},
\begin{align}
\frac{E(s)}{N}
=\frac{-1}{2}\left(\psi\!\left(\frac{A+1}{2}\right)
- \psi\!\left(\frac{A}{2}\right) \right)\nonumber \\
\label{eq:Ifinal1}
\end{align}
Putting the value of $A$:
\begin{equation*}
\frac{E(s)}{N}=\frac{-1}{2}\left(\psi\!\left(\frac{4s+1-2l+1+2}{4}\right)
- \psi\!\left(\frac{4s+1-2l+1}{4}\right)\right)
\label{eq:Ifinal2}
\end{equation*}
\begin{equation}
\implies\frac{E(s)}{N}=\frac{-1}{2}\left(\psi\!\left(\frac{4s+4-2l}{4}\right)
- \psi\!\left(\frac{4s+2-2l}{4}\right)\right)
\label{eq:Ifinal3}
\end{equation}
Reintroducing the summation in the integral \eqref{eq:Idef} admits the closed form
\begin{equation}
\frac{E(s)}{N}=\frac{-1}{2}\sum_{l=1}^{2s}\left(\psi\!\left(\frac{4s+4-2l}{4}\right)
     - \psi\!\left(\frac{4s+2-2l}{2}\right)\right)
     \label{eq:final1}
\end{equation}
Now evaluating different terms of the summation in \eqref{eq:final1}. Hence, for $l=1$:
\begin{equation*}
    \frac{-1}{2}\left(\psi\!\left(s+\frac{1}{2}\right)
     - \psi\!\left(\frac{s}{2}\right)\right)
\end{equation*}
For, $l=2$:
\begin{equation*}
    \frac{-1}{2}\left(\psi\!\left(s\right)
     - \psi\!\left(s-\frac{1}{2}\right)\right)
\end{equation*}
So the summation in \eqref{eq:final1} is a telescopic sum, hence, 
\begin{equation*}
\frac{E(s)}{N}=\frac{-1}{2}\left(\psi\!\left(s+\frac{1}{2}\right)
     - \psi\!\left(\frac{s}{2}\right) \cdots - \psi(1)+\psi(1)-\psi\left(\frac{1}{2}\right)\right)
\end{equation*}
\begin{equation}
\frac{E(s)}{N}=\frac{-1}{2}\left(\psi\!\left(s+\frac{1}{2}\right)
     - \psi\left(\frac{1}{2}\right)\right)
     \label{eq:final2}
\end{equation}
However, no step in the derivation of \eqref{eq:final2} relies on whether \(s\) is an integer or half-integer. Therefore, for integer \(s\),
\[\psi\left(s+\frac{1}{2}\right) = -\gamma - 2\ln{(2)} + \sum_{k=1}^{s} \frac{2}{2k-1} \] and 
\[\psi\left(\frac{1}{2}\right) = -\gamma -2\ln{(2)} \]
Where $\gamma = $ Euler-Mascheroni Constant. Now, from \eqref{eq:final2} one arrives at,
\begin{equation}
\frac{E(s)}{N}= -\sum_{k=1}^{s} \frac{1}{2k-1} 
    \label{eq:integer}
\end{equation}
Now for s: half integers, 
\[
\psi\left(s+\frac{1}{2}\right) = -\gamma  + \sum_{k=1}^{s-\frac{1}{2}} \frac{1}{k}
\]
Furthermore, upon inserting the known value of $\psi\!\left(\tfrac{1}{2}\right)$, the expression becomes:
\begin{equation}
\frac{E(s)}{N}= -\left(\ln{2}+ \sum_{k=1}^{s-\frac{1}{2}}\frac{1}{2k}\right)
    \label{eq:halfint}
\end{equation}
So, \eqref{eq:integer} and \eqref{eq:halfint} exactly match the proposed expression of Babujian \eqref{eq: babujan energy}.
\section*{Conclusion}
In this work, a compact analytical formula \eqref{eq:final2} has been proposed for the antiferromagnetic ground-state energy density of the isotropic Heisenberg spin-$s$ chain using the digamma function. This representation unifies the treatment of integer and half-integer spin, reproduces known results from the Takhtajan–Babujian solution, and bypasses the need for separate finite-series expansions. A promising direction for future work is to explore whether this digamma-based expression can be extended to anyonic spin chains governed by $SU(2)_k$ fusion rules~\cite{Feiguin2007,fendley1995rsos}. Such systems admit fractionalized spin degrees of freedom and exhibit rich topological and conformal structure. Developing an analogous exact formula in this setting could significantly advance the understanding of energy scaling, quantum group symmetries, and topological sectors in one-dimensional quantum systems.

\section*{Acknowledgements}
The author would like to thank Dr. Tibra Ali for his supervision and constant support throughout the work.

\end{document}